\documentclass[]{preprint}

\title{The Roles of Low-Noise Stations, Arrays and Ocean-Bottom Seismometers in Monitoring UK Offshore Seismicity associated with Subsurface Storage of Carbon Dioxide}

\author[1]{Dominik Strutz}
\author[1]{Andrew Curtis}

\leadauthor{Strutz and Curtis}
\shorttitle{Preprint submitted to International Journal of Greenhouse Gas Control}

\affil[1]{School of Geosciences, University of Edinburgh, Edinburgh EH93FE, UK, E-mail: dstrutz@ed.ac.uk}

\begin{document}
\maketitleabstract{
  \begin{summary}{Abstract}
      Effective seismic monitoring of subsurface carbon dioxide storage (SCS) sites is essential for managing risks posed by induced seismicity. This is particularly challenging in offshore environments, such as the Endurance license area in the North Sea, where the UK's permanent land-based seismometer network offers limited monitoring capability due to its distance from the expected locations of seismic events. A Bayesian experimental design framework is used to assess enhancements of the network with a low-noise onshore station located at around 1~km depth in Boulby mine, the onshore North York Moors Seismic Array, an optimally-located additional on-shore monitoring site, and ocean bottom seismometers (OBS). We quantify the expected information gain about seismic source locations and introduce a practical method to incorporate signal-to-noise dependent detectability and velocity model uncertainty. We show that the Boulby station or an onshore array primarily lower the detection threshold for small-magnitude events (M=0-2), but offer limited improvement in location accuracy. An optimally-located additional land-based seismometer or local array provides little additional benefit. OBS deployments yield significant improvements in location accuracy due to their proximity to potential seismicity. Optimised networks of two to three OBS stations are effective for Endurance, while three to five OBS stations offer robust monitoring across North Sea carbon storage licence areas off England's east coast. Velocity model uncertainty remains a key limiting factor for location precision across all configurations. We conclude that deploying OBS networks is the most promising strategy for enhancing microseismic monitoring capabilities at offshore SCS sites, though potentially more expensive.
  \end{summary}
}

\section{Introduction}

One of the key strategies in achieving zero net emissions of carbon dioxide to the atmosphere is the deployment of carbon capture and storage technologies. These involve capturing and injecting $CO_2$ into subsurface carbon dioxide storage (SCS) reservoirs. Such  reservoirs are usually below about 800m depth in order for the $CO_2$ to be in super-critical fluid state which makes direct monitoring of the migration of injected $CO_2$ difficult. However, stress changes associated with $CO_2$ migration tend to induce seismicity, so seismic events locations can be used to provide information about where $CO_2$ has migrated post-injection.

\begin{figure}
  \centering
  \includegraphics[width=0.5\textwidth]{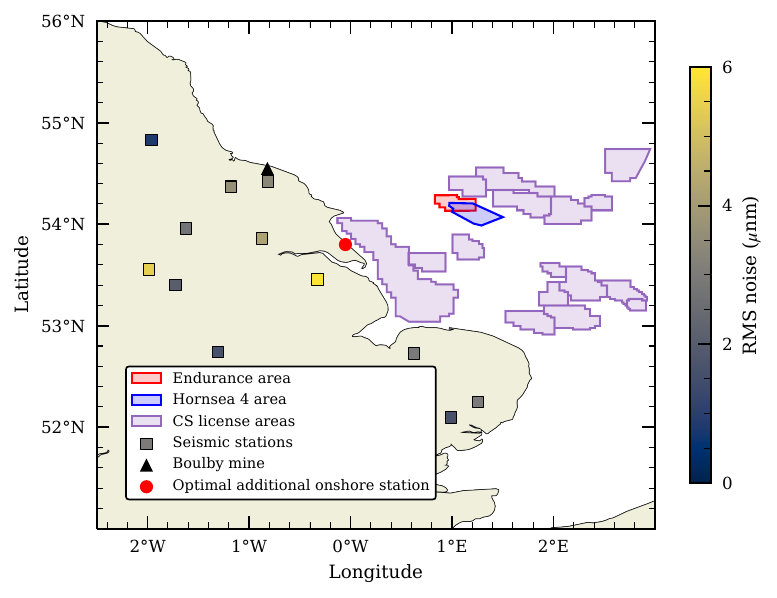}
  \caption{
    Geographic setup. The map shows the location of the Endurance and Hornsea 4 licence areas in the North Sea, as well as the location of the Boulby mine. The seismic stations used in this study are shown as rectangles, where the colour indicates the noise levels of the stations. See text for details about noise levels.
  }
  \label{fig:geographic_data}
\end{figure}

Effective monitoring of seismicity in such sites before, during, and after $CO_2$ injection is therefore essential for evaluating risk and ensuring safe operations. While induced seismicity is a concern in brittle crustal rocks that contain critically stressed faults \citep{Cerasi2018-ep, Oye2022-uu, Vilarrasa2019-th}, weak and poorly cemented sandstones in sedimentary basins are generally expected to deform slowly under tectonic forces, reducing the likelihood of faulting or fault reactivation \citep{Chadwick2004-xd}. Nevertheless, the potential for induced seismicity remains, and monitoring is necessary to ensure the safety of the operation, especially in areas where SCS sites are near offshore wind farms or other infrastructure. An example of such an area is the Endurance licence area in the North Sea, where the injection of $CO_2$ is planned near the Hornsea 4 offshore wind farm (see Figure \ref{fig:geographic_data}).

The UK's current seismometer network \citep{British-Geological-Survey1970-bd} is limited to a one-sided configuration with sensors only on the British mainland, resulting in a suboptimal configuration for monitoring offshore seismicity. This study investigates how adding low-noise seismic stations (e.g., in Boulby mine), seismic arrays (e.g., North York Moors Seismic Array), an optimally-located additional on-shore seismometer, and offshore seabed stations can enhance the monitoring of seismicity in the North Sea.

We use a Bayesian experimental design framework to determine the expected information gain (EIG) of different seismic network configurations. In our study, the EIG measures the expected reduction in uncertainty in the location of seismic events for any given seismic network configuration. We use the EIG to compare different configurations and provide examples of optimised networks with additional on-shore and OBS stations. We use a Bayesian design framework that accounts for the full uncertainty and non-linearity of the physics used to locate seismic sources. We show that the above land based interventions reduce the lowest detectable earthquake magnitudes across Endurance, without significantly improving the expected accuracy of event locations. The addition of three OBS stations in optimised locations would provide more information than any of the land based interventions tested, and would significantly improve confidence in the event locations. Five optimised OBS sensors may be sufficient to monitor Endurance and all of the neighbouring off-shore license areas.

\section{Methods}
We use Bayesian experimental design with a specific focus on seismic event monitoring. Experimental design is conducted before an experiment is carried out, and hence before data from the experiment are available. We therefore begin by introducing the prior information on event location and magnitude, and the data likelihood, which together describe the state of knowledge before we have collected any data. Thereafter we introduce the Bayesian experimental design framework which is used to determine the effectiveness of any particular design.
\subsection{Event Prior Information} \label{subsec:param_prior}

In this study we are interested in the source location of seismic events $\bm{m}$, which are the parameters of interest. Together with background seismic noise levels, the event location and magnitude ${M}$ governs the detectability of events. However, in this study we do not focus on estimating magnitudes since they are typically inferred directly from local moment magnitude relations \citep{Green2020-yx, Luckett2019-aw}. We therefore regard ${M}$ as a nuisance parameter. In other words, while both $\bm{m}$ and ${M}$ affect the observed data $\bm{d}$, we test and design optimal surveys to infer the source location $\bm{m}$ only.
\begin{figure}
  \centering
  \includegraphics[width=0.5\textwidth]{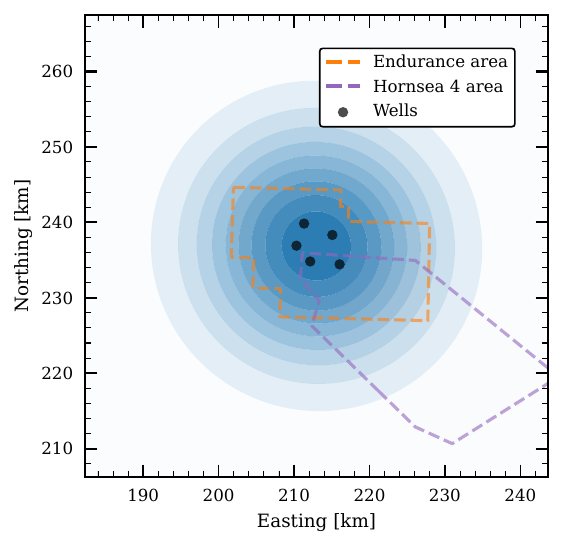}
  \caption{
    Prior distribution in event Easting and Northing (in local coordinates). Darker blue areas indicate higher probability density. The prior distribution is a mixture of five Gaussian distributions with a standard deviation of 10~km horizontally and 0.5~km vertically. The means of each Gaussian are located at the injection wells at a depth of 1020~m. The Endurance and Hornsea 4 licence areas are shown for reference (see Figure \ref{fig:geographic_data}).
  }
  \label{fig:prior_dist_EN}
\end{figure}

Prior information about the location of seismic events is typically based on the geological setting of the site and, in the case of SCS sites, seismicity is most likely to occur close to $CO_2$ injection point(s). We therefore model the prior probability distribution over event locations as a mixture of five Gaussian distributions, where the centre of each Gaussian is at the location of the injection wells, at a depth of ca. 1000~m which is the depth of the reservoir seal of the Endurance aquifer \citep{Sutherland2022-xj, Department-for-Business-Energy-Industrial-Strategy2023-gr, BP-Exploration-Operating-Company-Limited2023-jd}. The standard deviations are 10~km horizontally and 0.5~km vertically, chosen to be conservatively large to avoid introducing prior information that is too specific to any particular theory that forecasts locations of induced seismicity. The prior distribution in event Easting and Northing (in local coordinates) is shown in Figure~\ref{fig:prior_dist_EN}.
\begin{figure}
  \centering
  \includegraphics[width=0.5\textwidth]{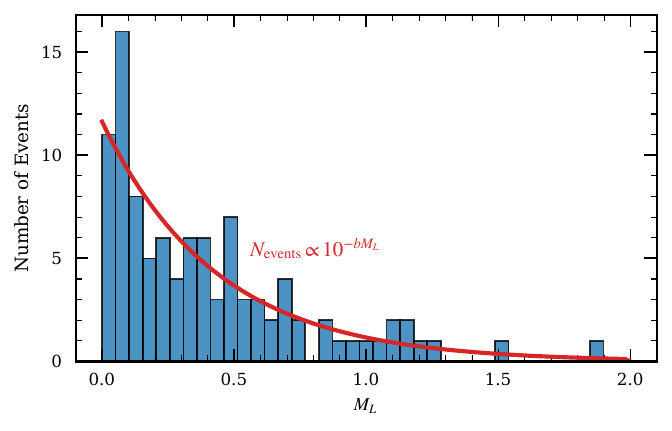}
  \caption{
    Distribution of event magnitudes. The distribution is a truncated Gutenberg-Richter distribution with a b-value of one. The histogram shows 100 realisations of the distribution. The red line indicates the analytical form of the distribution, scaled to the histogram.
  }
  \label{fig:nuisance_distribution}
\end{figure}

The distribution of event magnitudes is modelled as a truncated Gutenberg-Richter distribution \citep{Gutenberg1944-dd}, for which the frequency of an event with local magnitude $M_l$ is given by:
\begin{equation}
    f(M_l) = \frac{\exp\left[-\gamma \left( M_\text{l} - {M}_\text{lower}\right) \right]}
    {1 - \exp\left[-\gamma \left( {M}_\text{upper} - {M}_\text{lower}\right) \right]}+ {M}_\text{lower} \hspace{.5cm} \text{for } M_\text{l} \in \left[ {M}_\text{lower}, {M}_\text{upper} \right]
\end{equation}
where $\gamma=-b \cdot \log_{10}(e)$ is a scaling factor, $b$ is the b-value, which is chosen to be one here \citep{Kettlety2024-pl}, and $M_\text{lower}$ and ${M}_\text{upper}$ are the lower and upper magnitude bounds of interest. The distribution of event magnitude and a histogram of 100 realisations is shown in Figure~\ref{fig:nuisance_distribution}. Values of $M_\text{lower}$ and ${M}_\text{upper}$ will be assigned a variety of values in this study, which range from -1 to 3, spanning the range of magnitudes that are expected to occur within the detectable range of a seismic network in an offshore SCS setting. Small events are exponentially more likely to occur than large events and therefore dominate the analysis within any given magnitude range.

\subsection{Data Likelihood} \label{subsec:data_likelihood}

The data likelihood $p(\bm{d} \, | \, \bm{m},  M, \bm{\xi})$ describes the data $\bm{d}$ (a vector of P- and S-wave differential arrival times $d_{(i)} = t_{(i)}^{\text{S}} - t_{(i)}^{\text{P}}$, where $t_{(i)}^{\text{P}}$ and $t_{(i)}^{\text{S}}$ are the P- and S-wave arrival times at each of $N_\text{rec}$ receivers) that we expect to observe using a seismic network design described by the design vector $\bm{\xi}$, given a seismic event at location $\bm{m}$ with magnitude ${M}$. The use of arrival time differences allows the event origin time to be ignored in this study.

The data likelihood is assumed to be a multivariate Gaussian, with mean equal to the data predicted using a forward model described below, and a diagonal covariance matrix with appropriate entries for the expected uncertainties. This neglects correlations between the differential arrival times at different stations, but since our stations are generally geographically spread out, and since arrival time correlations appear to have a limited effect on the relative quality of different seismic network configurations \citep{Callahan2024-vu}, this is reasonable.
\begin{table}
  \centering
  \caption{Layered velocity model used for P-wave arrival time calculations.}
  \label{tab:velocity_model}
  \begin{tabular}{cc}
    \hline
    Depth (km) & $V_\text{p}$ (km/s) \\
    \hline
    0.00       & 4.00                \\
    2.52       & 5.90                \\
    7.55       & 6.45                \\
    18.87      & 7.00                \\
    34.15      & 8.00                \\
    \hline
  \end{tabular}
\end{table}

P-wave and S-wave arrivals are computed by ray-tracing through a layered medium using the Pyrocko software package \citep{Heimann2017-nf}. The velocity model used is the one employed by the British Geological Survey for earthquake localisation in the North Sea (refer to Table \ref{tab:velocity_model}). Calculated differential arrival times represent the mean of the data likelihood, and to account for the uncertainty in the data we model the variance of the likelihood as:
\begin{align}
  \sigma^2 & = \sigma_\text{pick}^2(\bm{m}, {M}, \bm{\xi}) + \sigma_\text{vel}^2(\bm{m}, \bm{\xi})                                                                  \\
           & =  \sigma_\text{pick (P-wave)}^2(\bm{m}, {M}, \bm{\xi}) + \sigma_\text{pick (S-wave)}^2(\bm{m}, {M}, \bm{\xi}) + \sigma_\text{vel}^2(\bm{m}, \bm{\xi})
\end{align}
Here, $\sigma_\text{pick}$ represents picking uncertainty, and $\sigma_\text{vel}$ represents the uncertainty due to our limited knowledge of the velocity model. In inverse problems, the velocity model term is often deterministic (and fixed) since there is only one realisation of the subsurface. However, in the context of experimental design, no data has yet been collected, and the effect of the velocity model on the expected data can be modelled as a random variable.
\begin{figure}
  \centering
  \includegraphics[width=0.5\textwidth]{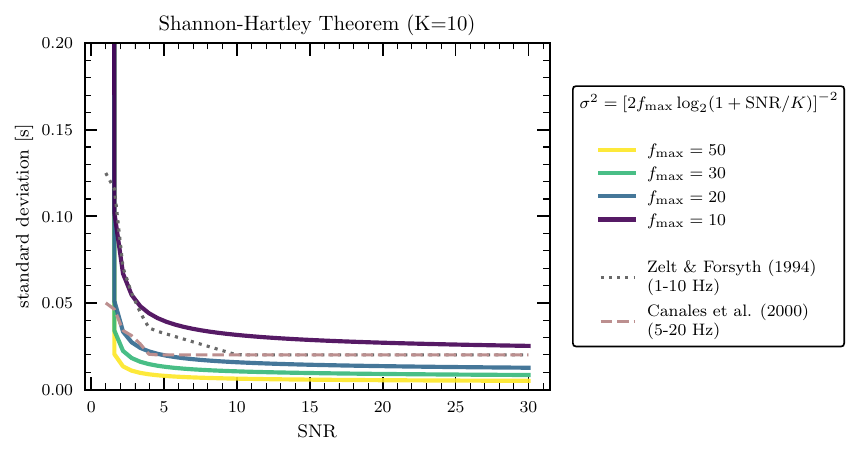}
  \caption{
    Picking uncertainty standard deviation in seconds as a function of the signal-to-noise ratio for different maximum frequencies. The empirical factor $K$ is set to 10. The empirical estimates of the picking uncertainty by \citet{Zelt1994-zf} and \citet{Pablo-Canales2000-tj} are shown for comparison.
  }
  \label{fig:shannon_hartley_theorem}
\end{figure}

The main factors governing picking uncertainty are the signal-to-noise ratio and the maximum frequency of the seismograms, and the picking method used \citep{Abakumov2020-xe, Fuggi2024-pu, Zhang2020-ga}. The picking uncertainty is assumed to be normally distributed with zero mean and a standard deviation of $\sigma_\text{pick}(\bm{m}, {M}, \bm{\xi})$. In this study, we follow the approach of \citet{Fuggi2024-pu} and \citet{Aki1976-zf}, and quantify the picking uncertainty using information theory (specifically the Shannon-Hartley theorem) as
\begin{equation}
    \sigma_\text{pick} (\bm{m}, {M}, \bm{\xi})^2 = \left[
      \log_2 \left( 1 + \frac{\text{SNR}(\bm{m}, {M}, \bm{\xi})}{K} \right) \, 2 \, f_\text{max}
      \right]^{-2}  \label{eq:Shannon_Hartley}
\end{equation}
\sloppy where $\text{SNR}(\bm{m}, {M}, \bm{\xi}) = 20 \log_{10} \left( A_\text{signal}(\bm{m}, {M}, \bm{\xi}) \,  /  \, A_\text{noise}(\bm{\xi}) \right)$ is the signal-to-noise ratio of the signal amplitude $A_\text{signal}(\bm{m}, {M}, \bm{\xi})$ and the noise amplitude $A_\text{noise}(\bm{\xi})$, and $f_\text{max}$ is the maximum frequency of the recorded signal. A similar approach has also been used by \citet{Keil2023-fz}. Figure \ref{fig:shannon_hartley_theorem} shows the picking uncertainty as a function of the signal-to-noise ratio for different values of $f_\text{max}$. The empirical factor $K=10$ is chosen differently from \citet{Fuggi2024-pu} and \citet{Aki1976-zf} (using K=10 instead of K=20) to align with the empirical estimates of the picking uncertainty by \citet{Zelt1994-zf} and \citet{Pablo-Canales2000-tj} (see Figure \ref{fig:shannon_hartley_theorem}). This increase in precision (decreased $K$) might be explained by technological advances in seismology since the 1970s, when the original formula was derived. In light of the ever-improving quality of machine learning phase pickers \citep[e.\,g.,\,][]{Zhu2018-lf}, we believe that the change in $K$ might even be conservative. It should be noted that some studies have suggested that the picking uncertainty is not normally distributed \citep{Diehl2009-nu, Abakumov2020-xe, Lomax2014-wr}, but since modelling the picking uncertainty as a normal distribution is common practice in seismology, we use that approach.

In this study we set the maximum frequency to 30~Hz, in line with \citet{Green2020-yx}, resulting in a sharp increase in picking uncertainty for SNR ratios close to one and a near constant uncertainty for SNR ratios above five. The main effect of the picking uncertainty is the sharp increase in uncertainty for SNR ratios close to one, a characteristic largely independent of the maximum frequency. If more information on the source fault model is available (such as stress drop and dislocation surface), $f_\text{max}$ can be calculated using, for example, Brune's model \citep{Brune1970-ny}.

\begin{figure}
  \centering
  \includegraphics[width=0.5\textwidth]{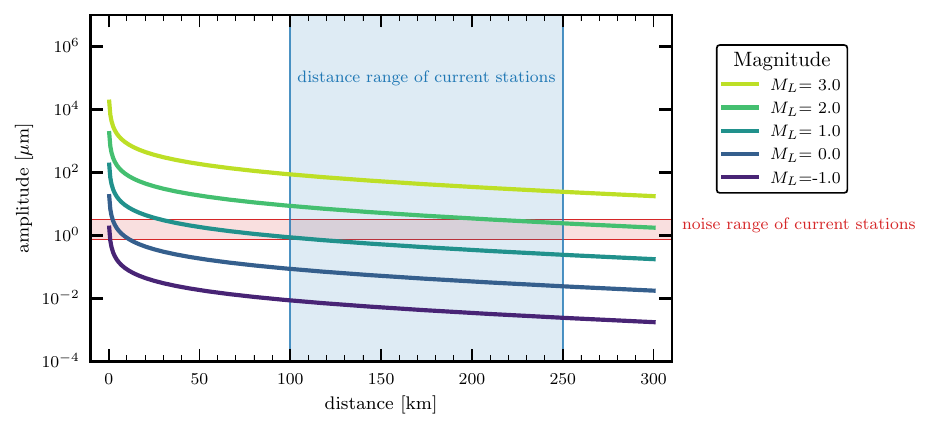}
  \caption{
    Signal amplitude as a function of epicentral distance for different local magnitudes. The signal amplitude represents the maximum amplitude of a P-wave measured in $\mu$m on a short-period Wood-Anderson seismometer. The signal amplitude is calculated using the local magnitude scale employed by the BGS \citep{Luckett2019-aw}. Noise levels (vertical component) and epicentral distance ranges of the stations and events used in this study are shown as shaded areas.
  }
  \label{fig:signal_amplitude}
\end{figure}
With the maximum frequency fixed, we need to find relations to model $A_\text{signal}(\bm{m}, {M}, \bm{\xi})$ and $A_\text{noise}(\bm{\xi})$ as functions of the source location, the event magnitude and the experimental design. Since we use differential arrival times, we need to model the amplitude of both the P- and S-wave signals. For the P-wave signal amplitude, we use the scale of \citet{Green2020-yx}
\begin{equation}
    {M} = \log_{10}\left(A^{\text{P-wave}}_\text{signal}\right) + 0.86 \log_{10} \left( r \right) + 0.0014 r - 0.95
  \label{eq:local_magnitude_pwave}
\end{equation}
and for the S-wave signal amplitude, we use the local magnitude scale routinely used by the BGS \citep{Luckett2019-aw}
\begin{equation}
    {M} = \log_{10}\left(A^{\text{S-wave}}_\text{signal}\right) + 1.11 \log_{10} \left( r \right) + 0.00189 r - 1.16 \exp^{-0.2 r} - 2.09
  \label{eq:local_magnitude_swave}
\end{equation}
where $r$ is the epicentral distance in km, and the signal amplitudes $A^{\text{P-wave}}_\text{signal}$ and $A^{\text{S-wave}}_\text{signal}$ are the maximum amplitudes of a P- or S-wave given in $\mu$m on a short-period Wood-Anderson seismometer \citep{Luckett2019-aw}, respectively.

The Equations \ref{eq:local_magnitude_pwave} and \ref{eq:local_magnitude_swave} allow us to convert the event magnitudes to signal amplitudes (see Figure \ref{fig:signal_amplitude}). To model the noise amplitude, we assume that the noise level of a station is constant over time, and calculate the power spectral density (PSD) of the noise for the vertical and horizontal components of each station over the last two years (excluding any station with significant gaps in the data) using the method of \citet{McNamara2004-gp}. We convert the PSD to the root-mean-square (RMS) amplitude of the noise in the frequency band of interest (10~Hz to 30~Hz) following \citet{Bormann2013-qx} as follows:
\begin{align}
  \text{RMS}_\text{noise} = \sqrt{2} \cdot \sqrt{\cdot \mathrm{PSD}_\text{mean} \cdot \left( f_\text{max} - f_\text{min} \right)} \label{eq:noise_amplitude}
\end{align}
where $f_\text{min}$ and $f_\text{max}$ are the lower and upper frequency bounds of the frequency band of interest, and $\mathrm{PSD}_\text{mean}$ is the mean PSD of the noise in the frequency band of interest. By multiplying the $\text{RMS}_\text{noise}$ with $\sqrt{2}$, we can convert the RMS amplitude to signal amplitude $A_\text{noise} = \sqrt{2} \cdot \text{RMS}_\text{noise}$, assuming a sinusoidal signal. The noise levels of the stations used in this study are shown in Figure \ref{fig:geographic_data}.

By comparing the fixed noise levels to the signal amplitudes (Equations \ref{eq:local_magnitude_pwave} and \ref{eq:local_magnitude_swave}), we can estimate the signal-to-noise ratio (SNR) for different event magnitudes and distances. We find that for the noise level and distance ranges considered in this study, the minimum detectable magnitude is around 1.0-2.0 (see Figure \ref{fig:signal_amplitude}). This is slightly more optimistic than the thresholds reported by the BGS \citep{Galloway2024-tq}, which assume a constant noise level of 10~$\mu$m (in a different frequency band than the one used here) for all stations. This difference should be taken into account when interpreting the results of this study. Considering that the BGS report \citep{Galloway2024-tq} shows detected events with magnitudes lower than their reported threshold, we believe that the detection threshold derived from the SNR used in this study is reasonable.
\begin{figure}
  \centering
  \includegraphics[width=0.5\textwidth]{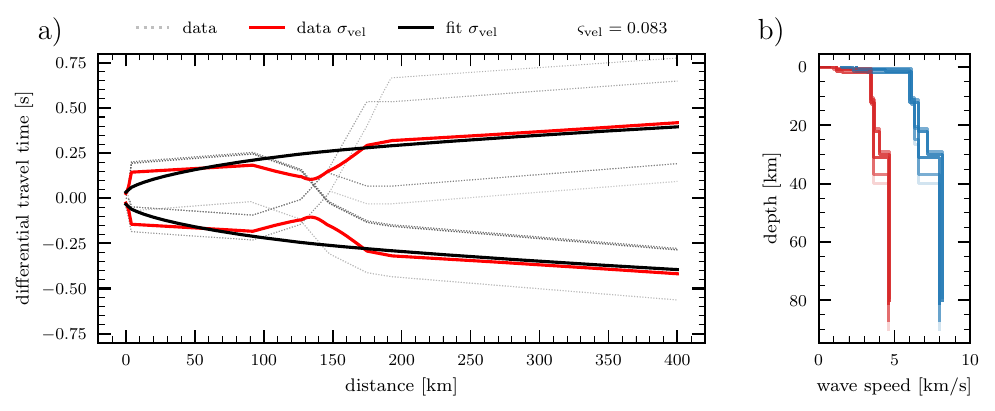}
  \caption{
    Panel (a) shows differential travel times for different velocity models as a function of epicentral distance in grey, the calculated standard deviation of the travel-time perturbations in red, and the fitted square root function in black. Panel (b) shows the P-wave (blue) and S-wave (red) velocity models used as a function of depth. These models are taken from CRUST 1.0 \citep{Laske2012-qy} within the area of interest.
  }
  \label{fig:velocity_std}
\end{figure}

To model $\sigma_\text{vel}^2(\bm{m}, \bm{\xi})$, we scale the differential travel-time $t$ by a factor $\varsigma_\text{vel}$, determined by the velocity model uncertainty, as follows:
\begin{align}
  \sigma_\text{vel}^2(\bm{m}, \bm{\xi}) = t(\bm{m}, \bm{\xi}) \cdot \left( \varsigma_\text{vel} \right)^2
\end{align}
This approach is inspired by a Gaussian random walk model, where the mean squared distance (in our case, differential travel-time noise) from the reference point (the mean differential travel-time) is proportional to time (differential travel-time measurement). We estimate the term $\varsigma_\text{vel}$ by calculating the standard deviation of differential travel-time perturbations from the mean for a selection of different velocity models. This is done by forward modelling travel-times as a function of distance for 25 crustal models (taken from CRUST 1.0 \citep{Laske2012-qy}, some are identical) within a square region bounded by $2.0^\circ$W to $2.0^\circ$E and $52.0^\circ$N to $56.0^\circ$N. We then fit a square root function to the standard deviation of these perturbations. Figure \ref{fig:velocity_std} shows the standard deviation of the differential travel-times for the different velocity models, the fitted square root function, and the velocity models with their respective (normalised) differential travel-times.
\begin{figure}
  \centering
  \includegraphics[width=0.5\textwidth]{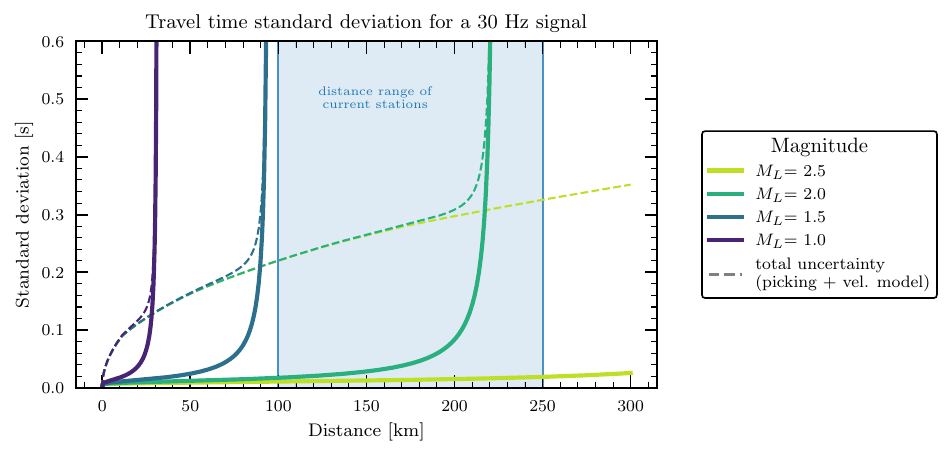}
  \caption{
    Travel time uncertainties of P-wave arrivals as a function of epicentral distance for different local magnitudes, at a fixed frequency of 30~Hz. The picking uncertainty is shown in colours ranging from blue to yellow indicating the magnitude $M$ in solid lines. The combined uncertainty is shown in dashed lines. The distance range of stations used in this study is indicated by the blue shaded area.
  }
  \label{fig:travel_time_st_ML}
\end{figure}

Because we use a horizontally-layered velocity model, we can express the differential travel time uncertainty as a function of epicentral distance, local magnitude and frequency, irrespective of lateral location. Figure \ref{fig:travel_time_st_ML} shows the picking and velocity model differential travel time uncertainty as a function of epicentral distance for different local magnitudes, at a fixed frequency of 30~Hz. The figure illustrates that the velocity model term dominates the total uncertainty up to the distance where the SNR approaches one, at which point the picking uncertainty increases sharply. This maximum effective distance is heavily dependent on the event magnitude.
\begin{figure}
  \centering
  \includegraphics[width=0.5\textwidth]{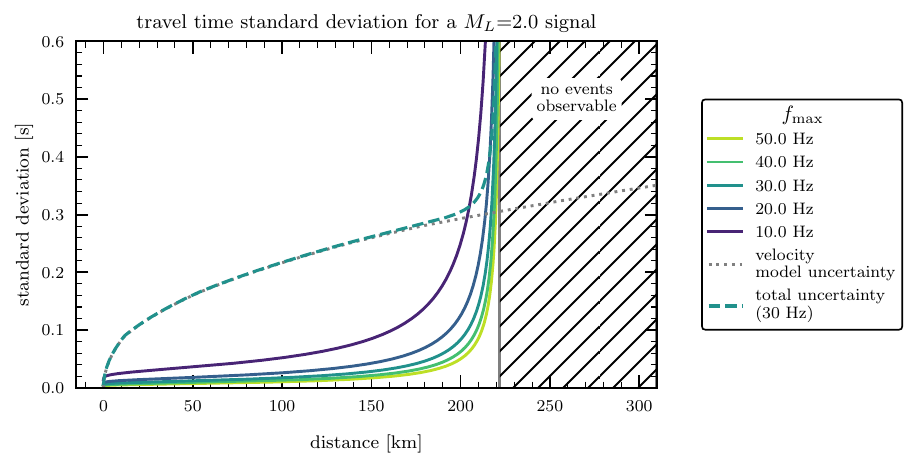}
  \caption{
    Travel time uncertainty as a function of epicentral distance for different frequencies, at a fixed local magnitude of ${M}=2.0$. The picking uncertainty is shown in colours ranging from blue to yellow indicating the frequency, whilst the velocity model term is shown in grey. The combined uncertainty for $30$~Hz is shown in dashed lines.
  }
  \label{fig:travel_time_st_freq}
\end{figure}

The effect of the maximum frequency on the uncertainty is shown in Figure \ref{fig:travel_time_st_freq}. While the maximum frequency does not change the maximum distance at which events are observable, a lower frequency will increase the picking uncertainty term. In addition, the sharp increase in uncertainty that occurs as the SNR approaches one is smoothed out and begins at smaller distances (Figure \ref{fig:travel_time_st_freq} shows a comparison with a fixed magnitude of ${M} = 2.0$). Only at frequencies of 1~Hz, the picking uncertainty term is roughly similar to the velocity model term, and would dominate if even lower frequencies were used. For the frequencies considered in this study, the velocity model term dominates the total uncertainty up to the distance where the SNR approaches one.

\subsection{Bayesian Experimental Design} \label{subsec:exp_design}
In sections \ref{subsec:param_prior} and \ref{subsec:data_likelihood}, we introduced the prior distributions for the parameters of interest $p(\mathbf{m})$, the nuisance parameters $p({M})$ and the data likelihood $p(\bm{d} \, | \, \bm{m},  M, \bm{\xi})$. After observing data $\bm{d}$, we can use this prior information to update our knowledge of the model parameters $\bm{m}$ and ${M}$ using Bayes' theorem:
\begin{equation}
  p(\mathbf{m}, {M} \, | \, \mathbf{d}, \bm{\xi}) = \frac{p(\mathbf{d} \, | \, \mathbf{m}, {M}, \bm{\xi}) \, p(\mathbf{m}) \, p({M})}
  {p(\mathbf{d} \, | \, \bm{\xi})}
\end{equation}
where $p(\mathbf{d} \, | \, \mathbf{m}, {M}, \bm{\xi})$ is the likelihood of recording data $\bm{d}$ given the event location $\bm{m}$ and magnitude ${M}$, and $p(\mathbf{d} \, | \, \bm{\xi}) = \int \int p(\mathbf{d} \, | \, \mathbf{m}, {M}, \bm{\xi}) \, p(\mathbf{m}) \, p({M}) \, d\mathbf{m} \, d{M}$ is called the evidence for data $\bm{d}$ given all assumptions made in the problem setup. Note the explicit dependence of $p(\mathbf{m} \, | \, \mathbf{d}, \bm{\xi})$ on the design: this is ignored in most geophysical inverse or inference problems, because the experimental design cannot be altered after data collection. All terms that involve observed data also depend on design \textemdash{} note the explicit $\xi$. To calculate the posterior distribution on only the parameters of interest, we marginalise over the nuisance parameters
\begin{equation}
  p(\mathbf{m} \, | \, \mathbf{d}, \bm{\xi}) = \int p(\mathbf{m}, {M} \, | \, \mathbf{d}, \bm{\xi}) \, d{M}
\end{equation}

The Bayesian framework provides a straightforward way to estimate the expected increase in information gained from collecting data for a specific experimental design $\bm{\xi}$. This expected information gain (EIG) is defined as the expected difference between the information content of the posterior distribution $p(\bm{m} \, | \, \bm{d}, \bm{\xi})$ and that of the prior distribution $p(\bm{m})$ \citep{Lindley1956-lu}:
\begin{equation}
  \operatorname{EIG}(\bm{\xi}) = \mathbb{E}_{p(\bm{d} | \bm{\xi})} \Bigl\{ \operatorname{I}[p(\bm{m} \, | \, \bm{d},\bm{\xi})] - \operatorname{I}[p({\mathbf{m}})] \Bigr\} \label{eqn:EIG_def_model}
\end{equation}
where $\operatorname{I}[p]$ denotes the Shannon information \citep{Shannon1948-of} of a distribution $p$ (see Appendix \ref{app:shannon_{(i)}nformation} for details), and $ \mathbb{E}_{p(\bm{d} \, | \,\bm{\xi})}$ represents the expectation taken over all possible data $\bm{d}$ that could be observed, given the experimental design $\bm{\xi}$ and the prior information. Intuitively, Equation \eqref{eqn:EIG_def_model} quantifies the expected reduction in uncertainty about the source locations $\bm{m}$, achieved by incorporating observations $\bm{d}$ from the experimental design $\bm{\xi}$, relative to the initial uncertainty based solely on the prior information. As actual data are unavailable during the design phase, this expectation is calculated by averaging over all plausible datasets $\bm{d}$ that might be observed. A selection of plausible datasets are generated by sampling potential earthquake locations $\bm{m}$ and magnitudes ${M}$ from their respective prior distributions, then simulating the corresponding data $\bm{d}$ using the likelihood function $p(\bm{d} \, | \, \bm{m}, {M}, \bm{\xi})$, which accounts for observational noise.

To evaluate the EIG in the form of Equation \eqref{eqn:EIG_def_model}, we would need to calculate the posterior distribution $p(\bm{m} \, | \, \bm{d}, \bm{\xi})$ for each possible observation $\bm{d}$, which is computationally expensive (unless we approximate the posterior distribution \citep{Foster2019-rx}). To make the problem tractable, we rearrange the EIG to depend explicitly on the evidence $p(\bm{d} \, | \, \bm{\xi})$ and the likelihood $p(\bm{d} \, | \, \bm{m}, \bm{\xi})$, instead of on the posterior distribution $p(\bm{m} \, | \, \bm{d}, \bm{\xi})$ and the prior distribution $p(\bm{m})$ (details in Appendix \ref{app:rearange_EIG}). The EIG can then be expressed as:
\begin{equation}
  \operatorname{EIG}(\xi) = \mathbb{E}_{p(\bm{m})} \Bigl\{ \operatorname{I}[p(\bm{d} \, | \, \bm{m}, \bm{\xi})] \Bigr\}- \operatorname{I}[p(\bm{d} \, | \,\bm{\xi})] \label{eqn:EIG_def_data}
\end{equation}
where the expectation is now over the prior distribution $p(\bm{m})$.

The EIG in this form can be estimated by the following Monte Carlo (MC) estimate:
\begin{equation}
  \operatorname{EIG}(\bm{\xi}) \approx \frac{1}{N} \sum_{i=1}^{N} \Bigl\{ \log[p(\bm{d}_{(i)} \, | \, \bm{m}_{(i)},\bm{\xi})] - \log[p(\bm{d}_{(i)} \, | \,\bm{\xi})] \Bigr\} \label{eqn:EIG_MC}
\end{equation}
where $\{\bm{m}_{(i)}, \bm{d}_{(i)}\}$ are drawn as follows: for each $i$, sample $\bm{m}_{(i)} \sim p(\bm{m})$ and $M_{(i)} \sim p(M)$ from the prior distributions; then sample $\bm{d}_{(i)} \sim p(\bm{d} \, | \, \bm{m}_{(i)}, {M}_{(i)}, \bm{\xi})$ from the likelihood. The samples $M_{(i)}$ are required to evaluate the likelihood, but do not appear explicitly in the MC estimate (Equation \eqref{eqn:EIG_MC}).

The evidence $p(\bm{d}_{(i)} \, | \,\bm{\xi})$ lacks a closed (explicit mathematical) form, but for each $i$ in Equation \eqref{eqn:EIG_MC} we can approximate it using another MC estimate \citep{Ryan2003-qp, Huan2013-nf}:
\begin{equation}
  p(\bm{d}_{(i)} \, | \,\bm{\xi}) \approx \frac{1}{N_\mathrm{evi}} \sum_{j=1}^{N_\mathrm{evi}} p(\bm{d}_{(i)} \, | \, \bm{m}_{(j)}, {M}_{(j)}, \bm{\xi}) \label{eqn:evidence_MC}
\end{equation}
We implement the computationally more efficient variation proposed by \citet{Huan2013-nf}, where the samples from the `inner loop' (Equation \eqref{eqn:evidence_MC}) are reused for the `  outer loop' (Equation \eqref{eqn:EIG_MC}). This approach reduces the number of required forward model calculations. Consequently, we set $N_\mathrm{evi} = N$.

Similarly, the data likelihood in the presence of nuisance parameters can be estimated as follows \citep{Feng2019-wp}:
\begin{equation}
  p(\bm{d}_{(i)} \, | \, \bm{m}_{(i)}, \bm{\xi}) \approx \frac{1}{N_\mathrm{like}} \sum_{k=1}^{N_\mathrm{like}} p(\bm{d}_{(i)} \, | \, \bm{m}_{(i)}, {M}_{(i,k)} , \bm{\xi}) \label{eqn:likelihood_MC}
\end{equation}
where ${M}_{(i,k)}$ are samples drawn from the nuisance parameter distribution $p({M})$. Substitution of the estimates from Equations \eqref{eqn:evidence_MC} and \eqref{eqn:likelihood_MC} into Equation \eqref{eqn:EIG_MC} yields a nested Monte Carlo estimate of the EIG:
\begin{align}
    \operatorname{EIG}_\text{NMC}(\bm{\xi}) \approx \frac{1}{N} \sum_{i=1}^{N} \left\{ \log \left[ \frac{1}{N_\mathrm{like}} \sum_{k=1}^{N_\mathrm{like}} p(\bm{d}_{(i)} \, | \, \bm{m}_{(i)}, {M}_{(i,k)} , \bm{\xi}) \right] - \log \left[ \frac{1}{N_\mathrm{evi}} \sum_{j=1}^{N_\mathrm{evi}} p(\bm{d}_{(i)} \, | \, \bm{m}_{(j)}, {M}_{(j)}, \bm{\xi}) \right] \right\}  \label{eqn:EIG_MC_nested}
\end{align}
While calculating the EIG in this form can be computationally more expensive than other more sophisticated methods \citep{Foster2019-rx, Feng2019-wp,Du2024-jx} or approximate methods \citep{Coles2011-ks,Long2013-uj}, it is straightforward to implement, asymptotically unbiased (it will converge to the true value with increasing number of samples), and serves as the standard benchmark for more sophisticated methods \citep{Foster2019-rx, Feng2019-wp, Goda2020-xj, Du2024-jx}. Since this study focuses primarily on analysing the information content of predefined designs, rather than searching through numerous designs to find the optimal one, the relatively high computational cost is acceptable. In this study, we use $N=N_\mathrm{evi}=10,000$ and $N_\mathrm{like}=500$ for the nested Monte Carlo estimate of the EIG (see Appendix \ref{app:convergence_tests} for convergence tests).

\section{Results}
We now present several ways to supplement the current UK seismic network and analyse how these would affect the expected information gain about seismic source locations across a range of event magnitudes. The expected information gain (EIG) measures network quality, over a large number of possible source locations (according to the prior distribution) rather than just one or a small number of events. While the EIG could be used directly to compare different designs, its value in units [nats] remains unintuitive and difficult to interpret. We therefore present an additional way to interpret the results, as follows.

If we approximate the posterior distribution by a multivariate isotropic Gaussian (with identical standard deviation $\sigma_\text{post}$ for all parameters and no inter-parameter correlations), we can use the expected posterior information
\begin{equation}
  \bar{\operatorname{I}}_\text{post} = \mathbb{E}_{p(\bm{m})}\left[ \operatorname{I}_\text{post} \right] = \operatorname{EIG} - \operatorname{I}_\text{prior}
\end{equation}
to estimate the expected posterior standard deviation
\begin{equation}
  \bar{\sigma}_\text{post}^2 = \exp{
    \frac{- \bar{\operatorname{I}}_\text{post}}{3} - \frac{1}{2} \left( 1 + \log(2 \pi) \right)} \label{eqn:post_std}
\end{equation}
where we have used the analytic expression for the Shannon information in a multivariate Gaussian. This expected standard deviation offers a more intuitive metric (in units of distance) for comparing different designs and analysing the expected uncertainty of the results. Due to the isotropic assumption, $\bar{\sigma}_\text{post}$ will underestimate horizontal uncertainties while overestimating vertical uncertainties, since the prior distribution constrains the vertical component more than the horizontal component. Throughout this study, we provide both measures EIG and $\bar{\sigma}_\text{post}$, and plots are scaled linearly with the expected standard deviation, which results in a non-linear scaling of the EIG in most plots.

\subsection{Low Noise Onshore Station \textemdash Boulby}\label{subsec:noise_boulby}
A low noise onshore seismic station might provide additional accuracy in source location estimates compared to the current UK seismic network. We first use the Boulby mine as a representative onshore location, as its depth and existing scientific infrastructure make it an ideal candidate. At 1100 - 1400~m below the surface, its geometry might also offer additional information compared to the current seismic network which is confined to the Earth's surface. The mine's proximity to the North York Moors Seismic Array (NYMA), which has been proposed for continuous monitoring of reservoir response to $CO_2$ injection in the North Sea \citep{Kettlety2024-qh}, implies that our results are relevant to both locations.

\begin{figure}
  \centering
  \includegraphics[width=0.5\textwidth]{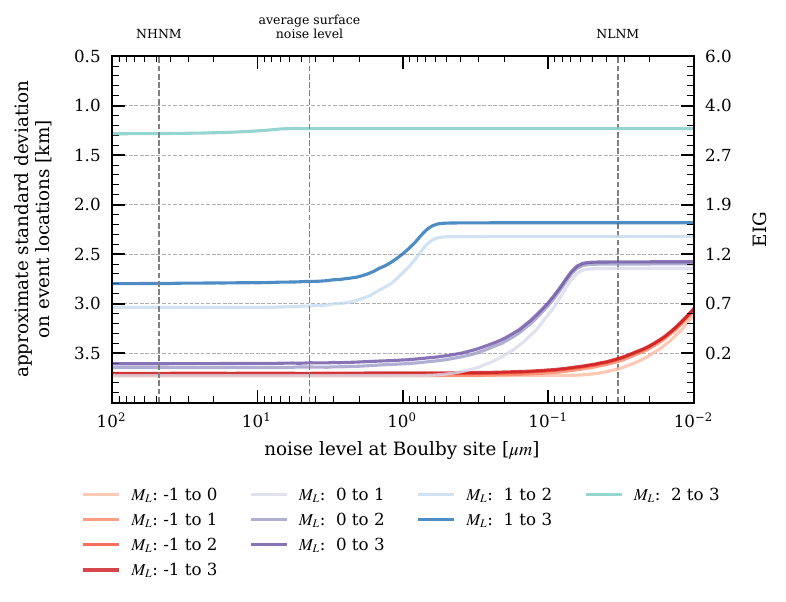}
  \caption{
    EIG and approximate posterior standard deviation as a function of noise level at Boulby (in $\mu$m) for different magnitude ranges. The EIG is calculated using the nested Monte Carlo estimate (Equation \eqref{eqn:EIG_MC_nested}) with $N=N_\mathrm{evi}=10,000$ and $N_\mathrm{like}=500$. The approximate posterior standard deviation in Equation \eqref{eqn:post_std} is derived from the expected information gain and the analytic expression for information in a multivariate Gaussian. The noise levels for an average UK surface station, the New High and Low Noise Models (NHNM, NLNM) are shown as reference with dashed lines. Note that station noise level decreases to the right of the plot, and expected event uncertainty increases downwards.
  }
  \label{fig:EIG_boulby_fixed_vel}
\end{figure}
Since the range of event magnitudes determines which events can be observed, we analyse the EIG as a function of noise level at Boulby (in $\mu m$) for ten different magnitude ranges between ${M}=-1$ and ${M}=3$ with results shown in Figure \ref{fig:EIG_boulby_fixed_vel}.

For the lowest magnitude ranges (starting at -1), the noise level at Boulby must be reduced substantially below $0.5 \cdot 10^{-2} ~\mu \text{m}$ to add any information beyond the permanent network. This noise range falls considerably below the new low noise model (NLNM) \citep{Peterson1993-ku}, which serves as a reference for low noise environments in a global context. While stations with noise levels below the NLNM are not impossible \citep{Berger2004-lu, Ringler2020-zu}, achieving this presents a significant challenge. The size of the Boulby mine might enable at local array to achieve such low noise levels \citep{Rost2002-mv}, but this could prove difficult to achieve close to an active mining operation.

Magnitude ranges from ${M}=\{0, 1\}$ to ${M}=\{1, 2, 3\}$ are where the Boulby station could provide the most benefit. Although the added expected information gain (compared to a very high noise level Boulby site) is slightly smaller than for lower magnitudes, this gain can be achieved at noise levels that are around an order of magnitude higher than the NLNM and around an order of magnitude lower than that of an average surface station in the UK. Given the depth of the mine, it is likely that the noise level can be reduced to this level, unless substantial noise is generated by the mining activities.

For the highest magnitude ranges (${M}\geq 2$), the Boulby site provides negligible additional information gain, even at very low noise levels. For these magnitudes, most events are already detected by the existing network, and the Boulby site offers no improvement in azimuthal coverage, nor event proximity needed to better constrain event depths. The primary benefit of the Boulby site for low-magnitude events therefore is its ability to detect small events that are unobserved by existing stations. Therefore, source location estimates for small events within the Endurance area are only possible if one can confirm that the event originates from the Endurance site rather than from elsewhere. Such verification could potentially be achieved through processing techniques which make use of a dense Boulby array \citep{Rost2002-mv}.
\subsubsection{Effect of Velocity Model Uncertainty}
\begin{figure}
  \centering
  \includegraphics[width=0.5\textwidth]{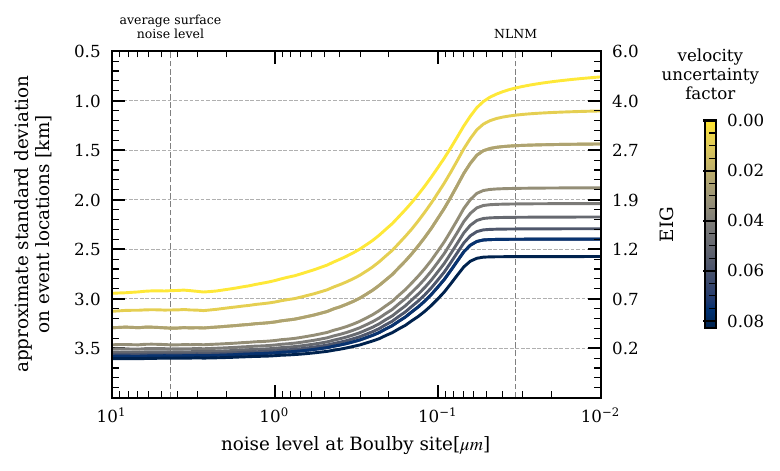}
  \caption{
    EIG and approximate posterior standard deviation as a function of noise level at Boulby (in $\mu$m) for a range of different velocity uncertainty factors. The EIG is calculated using the nested Monte Carlo estimate (Equation \eqref{eqn:EIG_MC_nested}) with $N=N_\mathrm{evi}=10,000$ and $N_\mathrm{like}=500$. The approximate posterior standard deviation is derived from the expected information gain and the analytic expression for information in a multivariate Gaussian (Equation \eqref{eqn:post_std}). The noise levels for an average UK surface station and the new low noise model (NLNM) are shown for reference with dashed lines. Note that station noise level decreases to the right of the plot, and expected event uncertainty increases downwards. 
  }
  \label{fig:eig_vel_uncertainty}
\end{figure}

For the previous analysis we have assumed a constant velocity uncertainty factor $\varsigma_\text{vel}$. The value of this factor (derived in section \ref{subsec:data_likelihood}) results in a data uncertainty that is dominated by the velocity uncertainty term until the signal-to-noise ratio approaches one. In this section we analyse the effect of lowering the velocity uncertainty on the expected information gain at the Boulby site, as a function of background noise level. This reduction in uncertainty might be achieved, for example, by performing seismic tomography for the velocity structure between Boulby and Endurance. We focus on the magnitude range ${M}=[0, 3]$ as this is where the Boulby site can provide the most benefit according to the analysis above. The results are shown in Figure \ref{fig:eig_vel_uncertainty}.

For high noise levels, decreasing the velocity uncertainty factor has only limited effect on the added EIG, since at these noise levels most events cannot be detected at the Boulby site or any other station. For lower noise levels, the velocity uncertainty has a substantial effect on the EIG, as at these levels every event could at least be detected at the Boulby station.

While in this synthetic example we can decrease the velocity uncertainty factor to zero, this is not possible in reality. Nevertheless, the results show that if the velocity uncertainty is reduced substantially, the approximate posterior standard deviation decreases by a factor of 2 to 3. This reduction will occur mostly in the horizontal components, since the vertical component cannot be well constrained due to the distance between the Boulby site and the event locations.

We also tested a grid of possible on-shore locations to identify the most informative site for a station (or array). The red station marked in Figure \ref{fig:geographic_data} proved optimal, situated at the closest onshore location to the Endurance site. However, results for that location did not differ substantially (EIG of 1.24 for a magnitude range of ${M}=[0, 3]$ at a noise level equal to the average onshore station) from those presented above for the Boulby site. Boulby achieves a similar EIG for the same magnitude range but at a slightly lower noise level, as it lies slightly further from the Endurance site. The information gain from this station therefore stems primarily from its ability to register more events, rather than from providing additional constraints on event locations.

\subsection{Onshore Seismic Arrays} \label{subsec:nyma_results}
Another method to improve the UK's seismic network is to add a (local) array of seismic stations, such as the North York Moors Seismic Array (NYMA). Array processing methods then allow calculation of the slowness vector, or the backazimuth and inclination of the incoming wavefield, which provides an alternative data type to the arrival time data used so far.
We found that even with an (unreasonably) sharp array response function, array data alone cannot provide substantial information beyond the travel-time data recorded by the current on-shore network (for details see Appendix \ref{app:nyma_results}). This remains true despite our over-generous assumption that the array can detect every event regardless of size, with uncertainty determined solely by the main lobe of the array response function. The results do not improve meaningfully even when placing the array at the closest onshore location to the Endurance site (see red station in Figure \ref{fig:geographic_data}, which is also the optimal location for a single additional station), rather than in the North York Moors.

To illustrate the measurement accuracy that would be needed for a seismic array to provide information beyond that provided by prior distribution, consider this simplified calculation: a Gaussian data error with $5^\circ$ standard deviation in backazimuth measurements corresponds to a location uncertainty of approximately 10~km in the orientation perpendicular to the backazimuth at a distance of 200~km. This closely matches the prior uncertainty of 10~km for each well. The value of $5^\circ$ is comparable to the backazimuth residuals reported by \citet{Jerkins2023-fb}. While array data can complement arrival time data, its uncertainty is too large to significantly constrain event locations beyond the prior. Similar to a low-noise station such as at the proposed Boulby site, the primary benefit of an onshore array is that the noise suppression of array methods allows detection of more (smaller) events \citep{Rost2002-mv,Jerkins2023-fb,Zarifi2023-eo}. Although not investigated here, offshore arrays have proven highly effective in monitoring potential SCS sites \citep{Jerkins2023-fb,Zarifi2023-eo}, at substantially increased cost.

\subsection{Ocean Bottom Sensors}

The information provided by on-shore stations is inherently limited, as stations are positioned far from expected event locations and often have restricted azimuthal coverage. Ocean-Bottom Seismometer (OBS) stations deployed on the seafloor offer an alternative to shore-based stations. Although OBS stations overcome many limitations of shore-based stations for seismic SCS monitoring, this flexibility entails higher costs and typically results in noisier seismic recordings \citep{Janiszewski2022-be}. In the following sections, we examine the additional benefits that OBS stations can provide when added to the current permanent seismic network of the UK.

\subsubsection{Single OBS Design \textemdash Endurance}

\begin{figure}
  \centering
  \includegraphics[width=0.5\textwidth]{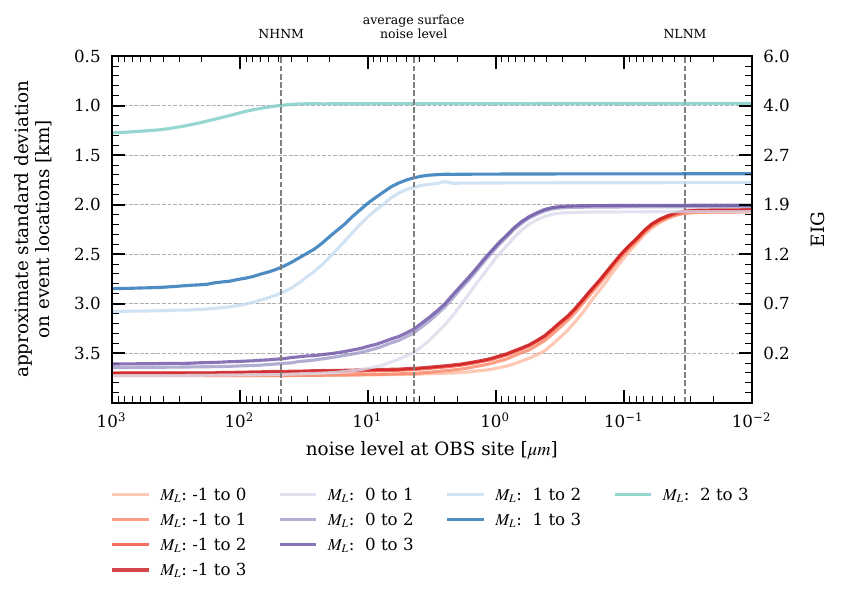}
  \caption{
    EIG and approximate posterior standard deviation as a function of noise level at an OBS station above the Endurance site (in $\mu$m) for different magnitude ranges. The EIG is calculated using the nested Monte Carlo estimate (Equation \eqref{eqn:EIG_MC_nested}) with $N=N_\mathrm{evi}=10,000$ and $N_\mathrm{like}=500$. The approximate posterior standard deviation is calculated using the expected information gain and the analytic expression for the information in a multivariate Gaussian in Equation \eqref{eqn:post_std}. The noise levels for an average UK surface station, and the New High and Low Noise Models (NHNM, NLNM) are shown as reference with dashed lines. Note that station noise level decreases to the right of the plot, and expected event uncertainty increases downwards. 
  }
  \label{fig:EIG_obs_fixed_vel}
\end{figure}

We first analyse the effect that the noise level of a single OBS station located directly above (at the center off of the set of wells) the Endurance site, in addition to the permanent stations, has on the EIG over several event magnitude ranges (similarly to section \ref{subsec:noise_boulby}). The results are shown in Figure \ref{fig:EIG_obs_fixed_vel}.

An OBS site with a noise level between the average UK surface station and the New High Noise Model (NHNM) \citep{Peterson1993-ku} can provide substantial additional information gain only for events with magnitude ${M} \geq 1$. At sufficiently low noise levels, this corresponds to a reduction of the expected posterior standard deviation from around $2.8$~km to $1.7$~km. While this improvement is significant, if we also wish to achieve a similar information gain for lower magnitude ranges then we require a noise level between the average UK surface station and the NLNM \citep{Peterson1993-ku}. At this noise level, the reduction in expected posterior standard deviation becomes even larger, dropping from $>3.5$~km to $2.0$~km. If such a noise level can be achieved, the OBS station would provide substantial additional information for all events with ${M} \geq -1$. As with the Boulby site, the benefit of adding a single OBS station is limited for events with ${M} \geq 2$, as these events are already well constrained by the existing network. For these events, the OBS station provides no meaningful additional information, even at low noise levels.

\begin{figure}
  \centering
  \includegraphics[width=0.5\textwidth]{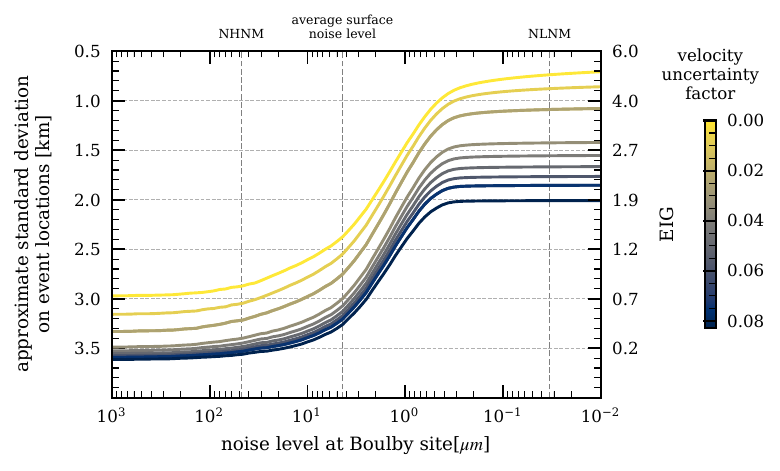}
  \caption{
    EIG and approximate posterior standard deviation as a function of noise level at an OBS station (in $\mu$m) for different velocity uncertainty factors. The EIG is calculated using the nested Monte Carlo estimate (Equation \eqref{eqn:EIG_MC_nested}) with $N=N_\mathrm{evi}=10,000$ and $N_\mathrm{like}=500$. The approximate posterior standard deviation is derived from the expected information gain and the analytic expression for information in a multivariate Gaussian (Equation \eqref{eqn:post_std}). The noise levels for an average UK surface station and the New High and Low Noise Models (NHNM, NLNM) are shown for reference with dashed lines. Note that station noise level decreases to the right of the plot, and expected event uncertainty increases downwards.
  }
  \label{fig:eig_vel_uncertainty_obs}
\end{figure}
We can also analyse the effect of velocity model uncertainty on the EIG at the OBS site, as we did for the Boulby site in section \ref{subsec:noise_boulby}. The results are shown in Figure \ref{fig:eig_vel_uncertainty_obs}. The general pattern is similar to that for the Boulby site, with the EIG increasing sigmoidally as the noise level decreases. However, the inflection point of this curve is shifted to lower noise levels, as the OBS station is substantially closer to the Endurance site than the Boulby site. While the additional distance of the Boulby station leads to a lower EIG compared to an OBS for large velocity uncertainty factors, for small velocity uncertainty factors the achievable EIG from a distant station is only slightly lower than that from a nearby station.

\subsubsection{Designs Consisting of Multiple OBS \textemdash Endurance}
\begin{figure}
  \centering
  \includegraphics[width=1.0\textwidth]{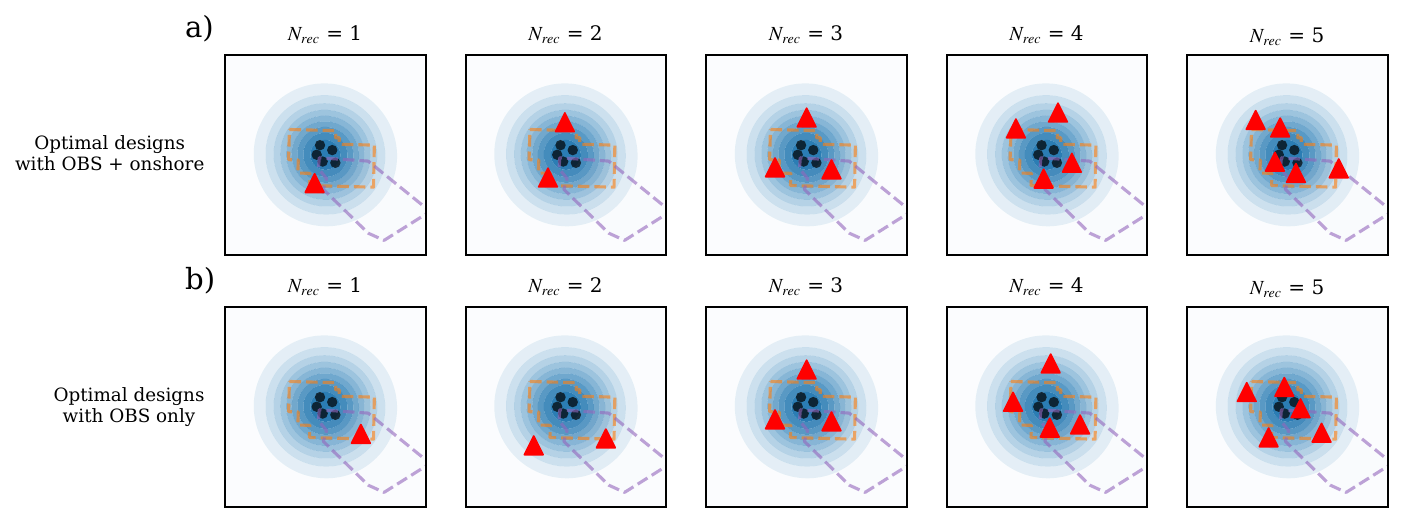}
  \caption{
    Optimal OBS locations for networks containing up to five OBS stations (red). The upper row shows the optimal OBS designs for a combined network of existing onshore stations and OBS stations, whilst the lower row displays the optimal designs for a standalone OBS network.
  }
  \label{fig:oed_optimise_OBS_optimal_designs_combined}
\end{figure}
Since the OBS station results are the most promising for the Endurance site, we now analyse the effect of adding multiple OBS stations to the network. Figure \ref{fig:oed_optimise_OBS_optimal_designs_combined} shows the optimal designs for networks of up to five OBS stations. For this analysis a magnitude range of ${M}=[0, 3]$ is used, as this is where OBS stations could provide the most benefit, if their noise levels are sufficiently low. These designs were optimised using Bayesian optimisation \citep{Mockus1975-js, Shahriari2016-an}, an efficient global-search algorithm. As the prior information in depth is stronger than in the horizontal components, the designs do not necessarily require a station directly above the injection points. If depth is of particular importance, the EIG algorithms can be adapted to treat the horizontal components of the event locations as nuisance parameters and thus focus more, or indeed solely on the depth component \citep{Strutz2023-pl, Feng2019-wp}.
\begin{figure}
  \centering
  \includegraphics[width=0.5\textwidth]{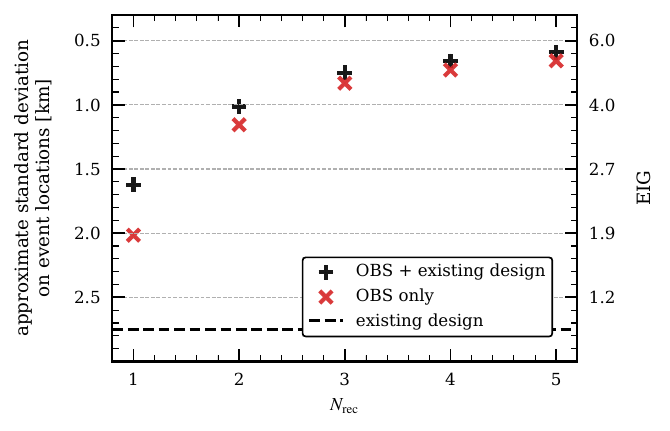}
  \caption{
    EIG and approximate posterior standard deviation as a function of the number of OBS stations for a network of combined OBS plus existing stations (black +), and for a stand-alone OBS network (red x), based on the optimal designs shown in Figure \ref{fig:oed_optimise_OBS_optimal_designs_combined}.
  }
  \label{fig:oed_optimise_OBS_eig_vs_N_rec}
\end{figure}

Figure \ref{fig:oed_optimise_OBS_eig_vs_N_rec} illustrates the performance of the designs presented in Figure \ref{fig:oed_optimise_OBS_optimal_designs_combined}. All OBS designs substantially outperform the baseline EIG of the existing network, with EIG increasing as the number of OBS stations grows. Onshore stations then provide limited additional information even if only one OBS station is deployed, and their contribution becomes negligible once more than two OBS stations are in place. For monitoring seismicity in the Endurance area specifically, there are diminishing returns when deploying more than three OBS stations.

\subsubsection{Designs Consisting of Multiple OBS \textemdash All Licences} \label{subsubsec:mul_obs_all_licences}
\begin{figure}
  \centering
  \includegraphics[width=1.0\textwidth]{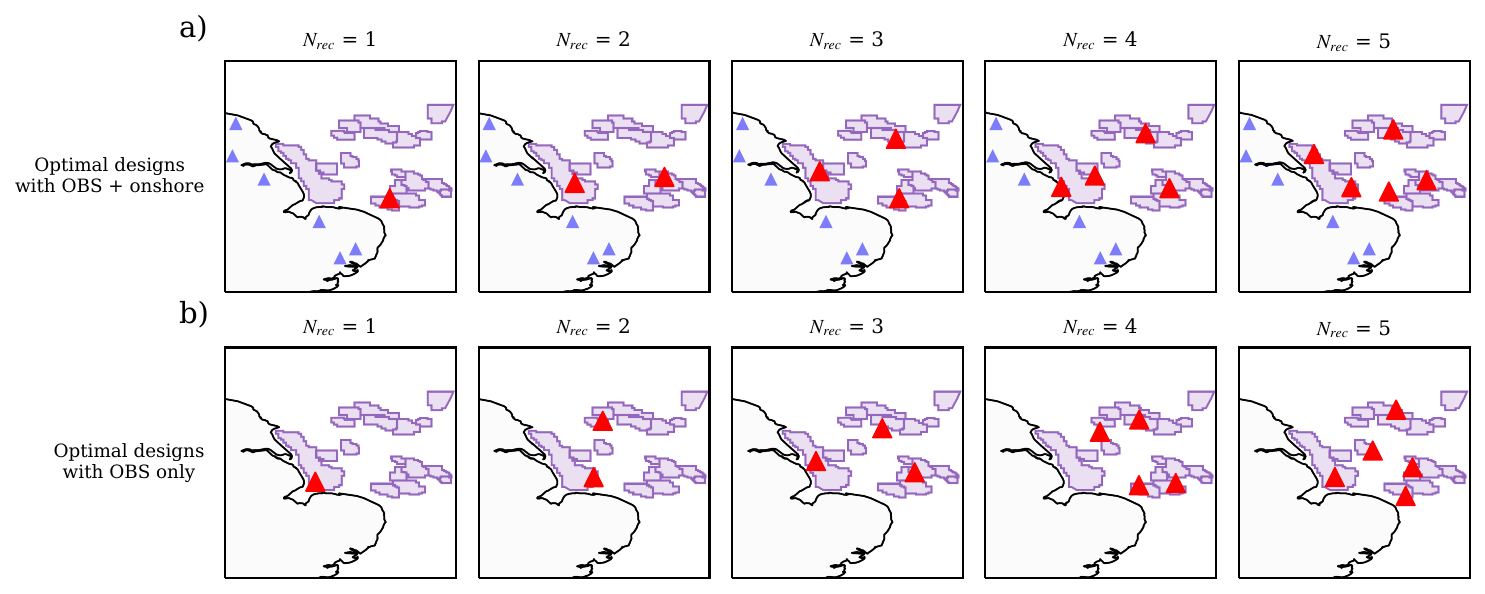}
  \caption{
    Optimal OBS locations for networks of up to five OBS stations for all potential SCS sites off the south-eastern coast of the UK. The upper row shows the optimal designs for a combined network of onshore (blue) and OBS stations (red), while the lower row displays the optimal designs for a standalone OBS network (red). UK coastline is shown in black.
  }
  \label{fig:oed_all_NS_optimise_OBS_position_combined}
\end{figure}

So far we have only considered the Endurance site, but the UK has several other potential SCS sites off its south-eastern coast. We now analyse the effect of a small network of OBS stations when added to the current UK seismic network, to monitor all of these sites. Using the same methodology as in the previous section, we employ a different prior distribution: a uniform distribution within the bounds of all potential SCS sites for the horizontal components, and a truncated Gaussian distribution for depth (mean: 1000~m, standard deviation: 500~m, truncated between 100~m and 2000~m). The resulting optimal OBS networks, both as combined and standalone configurations, are shown in Figure \ref{fig:oed_all_NS_optimise_OBS_position_combined}.
\begin{figure}
  \centering
  \includegraphics[width=0.5\textwidth]{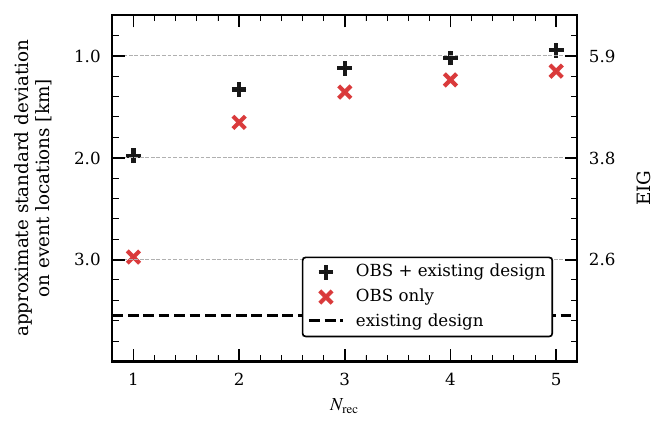}
  \caption{
    EIG and approximate posterior standard deviation as a function of the number of OBS stations for combined (black +) and standalone (red x) networks, based on the optimal designs shown in Figure \ref{fig:oed_all_NS_optimise_OBS_position_combined}.
  }
  \label{fig:oed_all_NS_optimise_OBS__eig_vs_N_rec}
\end{figure}
The performance of the designs shown in Figure \ref{fig:oed_all_NS_optimise_OBS_position_combined} is shown in Figure \ref{fig:oed_all_NS_optimise_OBS__eig_vs_N_rec}. A single standalone OBS sensor demonstrates similar performance to the existing UK seismic network. It is important to keep in mind that this is only true under the assumption that any seismicity recorded originated from one of the licences. Given the central location of the OBS station(s), polarity derived backazimuths could be used to test this if a three component OBS is deployed. Combining a single OBS with the on-shore network leads to a substantial increase in EIG. With three or more OBS stations, a standalone OBS network performs nearly as well as a combined network. For five or more stations, there are diminishing returns in adding additional stations.

Several optimal OBS locations for all licence areas (see Figure \ref{fig:oed_all_NS_optimise_OBS_position_combined}) appear close to the coast, which might permit similar performance to an OBS-only network through a mix of onshore and offshore stations. Given the high cost of OBS station deployment, a slight performance reduction from moving stations to the coast might be acceptable. It would also then be possible to directly optimise a mixed network of onshore and offshore stations.

\section{Discussion and Conclusion}
We employed a Bayesian experimental design framework to evaluate the expected information gain (EIG) from potential enhancements to the UK's seismic network for monitoring subsurface carbon storage (SCS) sites, specifically focusing on the Endurance area and broader North Sea licence areas. Our approach quantifies the expected reduction in uncertainty about seismic source locations, accounting for non-linearities and parameter uncertainties without relying on linearised approximations. The data likelihood model, which incorporated the signal-to-noise ratio (SNR) for event detectability and velocity model uncertainty for location precision, is central to our analysis.

Adding a single, low-noise onshore station, such as one at Boulby, can improve monitoring capabilities, particularly for events with magnitudes between $M=0$ and $M=2$. However, the EIG is limited for larger events ($M \geq 2$) since the existing network can already detect these events. The primary benefit stems from detecting smaller events missed by the current network, rather than substantially improving the location accuracy of already detected events. Achieving significant gains for magnitudes ${M} < 0$ requires exceptionally low noise levels, potentially below the NLNM, which is unlikely to be feasible in practice.

Our analysis highlights that uncertainty in the seismic velocity model is the dominant factor limiting the accuracy of source location estimates once an event has been detected (i.e., when SNR is sufficiently high). Reducing this uncertainty, perhaps through dedicated local seismic tomographic surveys, would yield substantial improvements in location accuracy, especially for networks capable of detecting events, such as those including low-noise stations or OBS. The velocity uncertainty estimate used here, derived from regional models, might be conservative for specific SCS sites where local subsurface imaging is often available.

Similarly to a single low-noise station, the main advantage of an onshore seismic array like NYMA, or even an optimally located array, lies in its potential for enhanced detection of small-magnitude events through array processing techniques that suppress noise. The directional information (slowness vectors) provided by the array, given typical uncertainties and distances, offers limited additional constraint on source locations compared to either arrival times recorded across the network or the prior information.

Ocean Bottom Seismometers (OBS) offer significant benefits. Deploying even one OBS near the target area provides substantial EIG, primarily due to its proximity, which increases signal amplitudes. An OBS with noise levels comparable to an average UK surface station can substantially improve monitoring for events ${M} \geq 1$. Capturing smaller events (${M} < 1$) requires lower noise levels, approaching the NLNM. Optimised networks of 2-3 OBS stations provide most of the achievable information gain for the Endurance site specifically. Beyond three stations, the returns diminish, and the contribution of the existing onshore network becomes marginal. For monitoring seismicity across all potential licence areas off the UK's south-eastern coast, a network of 3-5 OBS stations offers a robust solution. A standalone network of three OBS performs nearly as well as a combined network including both offshore and onshore stations. Adding more than five OBS yields progressively smaller improvements in EIG, at a level that may not justify the cost of deployment, since OBS are expensive even in the shallow North Sea.

In conclusion, the current UK seismic network provides adequate monitoring for larger magnitude events (${M} \geq 2$) near the Endurance site. Enhancing the capability to detect and locate smaller magnitude events necessitates additions to the network. The most effective strategies involve either deploying onshore stations/arrays with exceptionally low noise levels (primarily improving detection) or deploying Ocean Bottom Seismometers (improving both detection and location accuracy due to proximity). The latter provide significantly greater accuracy in the results, but probably at increased cost, producing a trade-off for decision-makers to consider. 

\section{Acknowledgements}

The implementations of the OED algorithms in this work would not have been possible without extensive use of open-source software. Not all of them have been included in the respective sections to ease readability. All the code was written in Python \citep{Van_Rossum2011-ep}, the libraries PyTorch \citep{Paszke2019-lv} and Zuko \citep{Rozet2023-zy} were used to process probability distributions and calculate the EIG. ObsPy \citep{Beyreuther2010-vm, Megies2011-nu, Krischer2015-ni} was used to process seismic data, and calculate the power spectral densities. NumPy \citep{Harris2020-nt} was used for general data processing and Matplotlib \citep{Hunter2007-jw} for plotting. The library \textit{Bayesian Optimization} \citep{Nogueira2014-vf} was used to find the optimal designs.

We want to thank Tom Kettlety for providing helpful information on the Boulby Mine and its current instrumentation, and Brian Baptie for providing noise levels for the UK seismic network used in an earlier version of this study.

This project has received funding from the European Union’s Horizon 2020 research and innovation programme under the Marie Skłodowska-Curie grant agreement No 955515 – SPIN ITN (www.spin-itn.eu)

\section{Declaration of generative AI and AI-assisted technologies in the writing process.}

During the preparation of this work the author(s) used LLMs in order to assist in improving the clarity and readability of the text. After using this tool/service, the author(s) reviewed and edited the content as needed and take(s) full responsibility for the content of the published article.

\bibliography{Strutz_Curtis_2025_ccs_monitoring.bib}

\appendix
\section{Shannon Information}\label{app:shannon_{(i)}nformation}

Shannon information \citep{Shannon1948-of} is a measure of information with beneficial properties (e.g., linear additivity of information from independent sources). The Shannon information $\operatorname{I}[\cdot]$ of an arbitrary continuous probability density function $p(x)$ is defined as
\begin{align}
  \begin{split}
    \operatorname{I} \left[p(x) \right] &= \mathbb{E}_{p(x)} \left[ \log _{b}\left(p(x) \right) \right] \\
    &= \int_{\mathcal{X}} p(x) \log _{b}(p(x)) d x
  \end{split}
  \label{eqn:information}
\end{align}
where $x\in \mathcal{X}$ is a random variable distributed according to $p(x)$ and $\mathbb{E}_{p(x)}$ is the expectation with respect to $p(x)$, which is defined by the right-most expression. Information is also often expressed through the entropy $\operatorname{H}[p(x)] = - \operatorname{I} \left[p(x) \right]$, where entropy $\operatorname{H}$ is defined as the negative of either expression on the right of Equation \eqref{eqn:information}.

\section{Rearranging the Expected Information Gain}\label{app:rearange_EIG}
\begin{align}
  \operatorname{EIG}(\bm{\xi}) & = \mathbb{E}_{p(\bm{d} | \bm{\xi})} \left[\operatorname{I}[p({M} \, | \, \bm{d},\bm{\xi})] - \operatorname{I}[p({{M}})] \right]                             \\
                     & = \mathbb{E}_{p(\bm{d}, {M} | \bm{\xi})} \left[ \log \frac{p({M} \, | \, \bm{d},\bm{\xi})}{p({{M}})} \right]                            \\
                     & = \mathbb{E}_{p(\bm{d}, {M} | \bm{\xi})} \left[ \log \frac{p({M}, \bm{d} \, | \, \bm{\xi})}{p({{M}}) p(\bm{d} \, | \,\bm{\xi})} \right] \\
                     & = \mathbb{E}_{p(\bm{d}, {M} | \bm{\xi})} \left[ \log \frac{p(\bm{d} \, | \, {M},\bm{\xi})}{p(\bm{d} \, | \, \bm{\xi})} \right]          \\
  \operatorname{EIG}(\bm{\xi}) & = \mathbb{E}_{p({M})}\left[ \operatorname{I}[p(\bm{d} \, | \, {M},\bm{\xi})] - \operatorname{I}[p(\bm{d} \, | \,\bm{\xi})] \right]
\end{align}

\section{Convergence Tests}\label{app:convergence_tests}
\begin{figure}
  \centering
  \includegraphics[width=0.5\textwidth]{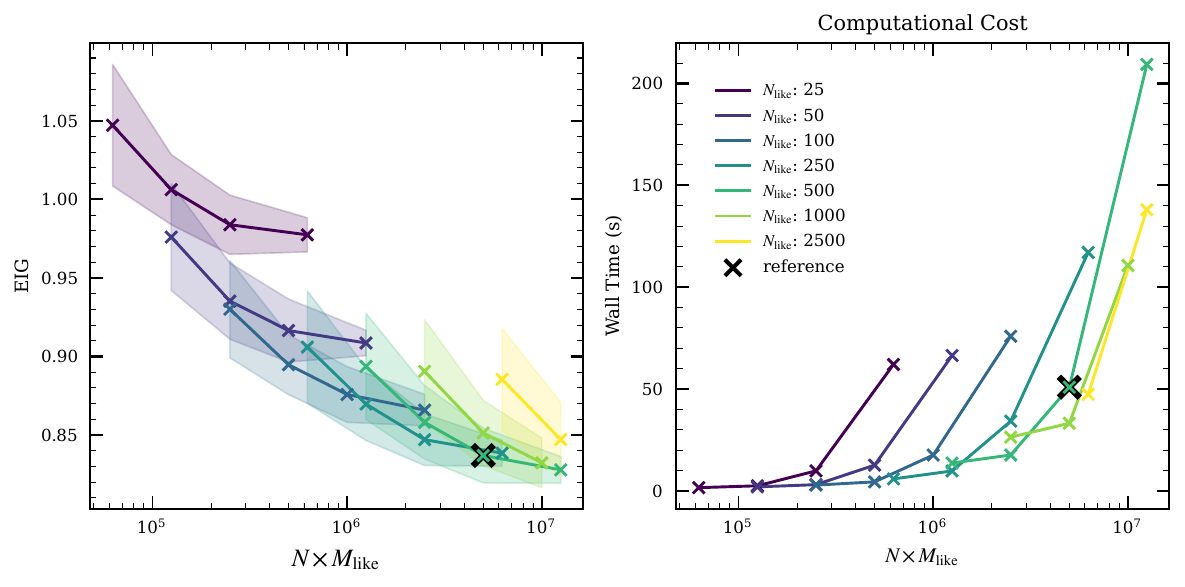}
  \caption{
    Convergence of the nested Monte Carlo EIG estimate (Equation \eqref{eqn:EIG_MC_nested}). The reference values are $N_\mathrm{evi}=10,000$ and $N_\mathrm{like}=500$. The left plot shows the convergence of the estimate as a function of the total number of samples ($N_\mathrm{evi} \times N_\mathrm{like}$) for different values of $N_\mathrm{like}$ and $N_\mathrm{evi}$. The line and shading indicate the mean and standard deviation of the EIG estimate (50 trials) respectively. The right plot shows the required wall time as a function of the total number of samples ($N_\mathrm{evi} \times N_\mathrm{like}$) for different values of $N_\mathrm{like}$ and $N_\mathrm{evi}$.
  }
  \label{fig:EIG_convergence_reuse}
\end{figure}
The choice of sample numbers $N$ and $N_\mathrm{like}$ in the nested Monte Carlo estimate of the EIG (Equation \eqref{eqn:EIG_MC_nested}) is crucial for both accuracy and computational cost. This section analyses the convergence of the EIG estimate for the current UK network plus a Boulby mine station. Figure \ref{fig:EIG_convergence_reuse} shows that at least $N_\mathrm{like}=500$ samples are needed to estimate the data likelihood accurately in the presence of nuisance parameters. With sufficiently large $N_\mathrm{like}$, increasing the number of outer loop samples $N$ primarily reduces variance in the EIG estimate. For this study, we use $N=N_\mathrm{evi}=10,000$ and $N_\mathrm{like}=500$ as a compromise between computational cost and accuracy. The accuracy reported in Figure \ref{fig:EIG_convergence_reuse} is most relevant for absolute results; for relative differences, results are more accurate when the same model and nuisance samples are used across comparisons.

\section{NYMA results}\label{app:nyma_results}
This section supplements section \ref{app:nyma_results} and provides additional information on the results of adding the NYMA array to the current UK seismic station network. Here we examine the array in isolation, where the only recorded data are the horizontal apparent slowness components, or equivalently, the backazimuth and inclination of the incoming wavefield.

\begin{figure*}
  \centering
  \includegraphics[width=1.0\textwidth]{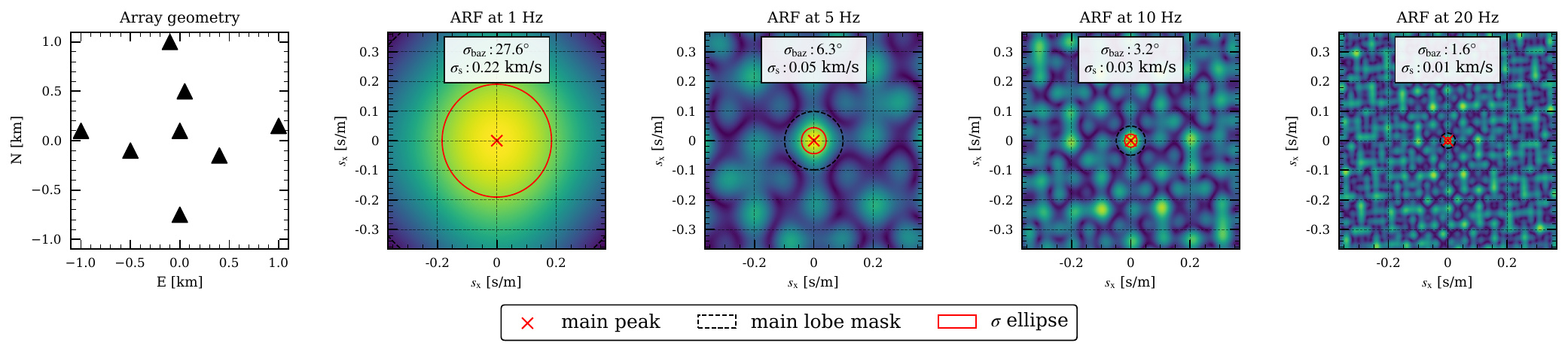}
  \caption{
    Array response function for the NYMA array. The left-most plot shows the geometry of the array, the other four plots show the array response function for frequencies from 1~Hz to 20~Hz. The backazimuth uncertainty is given for an inclination of 25$^\circ$ and a subsurface velocity of 3000~m/s.
  }
  \label{fig:array_arf_cov}
\end{figure*}
We use the array response function \citep{Rost2002-mv} to quantify the data likelihood of the array data. To simplify the problem, we assume that the array can detect every event regardless of its size, with uncertainty determined solely by the main lobe of the array response function. We then fit a Gaussian distribution to this main lobe to quantify the array data uncertainty. As this uncertainty in slowness is not always intuitive, we also report the approximate backazimuth uncertainty $\sigma_{\text{baz}}$ for an inclination of 25$^\circ$ and a subsurface velocity of 3000~m/s. The chosen inclination aligns with the expected angle of the incoming wavefield for a source at approximately 1~km depth and 200~km distance. The array response functions for frequencies from 1~Hz to 20~Hz, along with the slowness and backazimuth uncertainty, are shown in Figure \ref{fig:array_arf_cov}.

\begin{figure}
  \centering
  \includegraphics[width=0.5\textwidth]{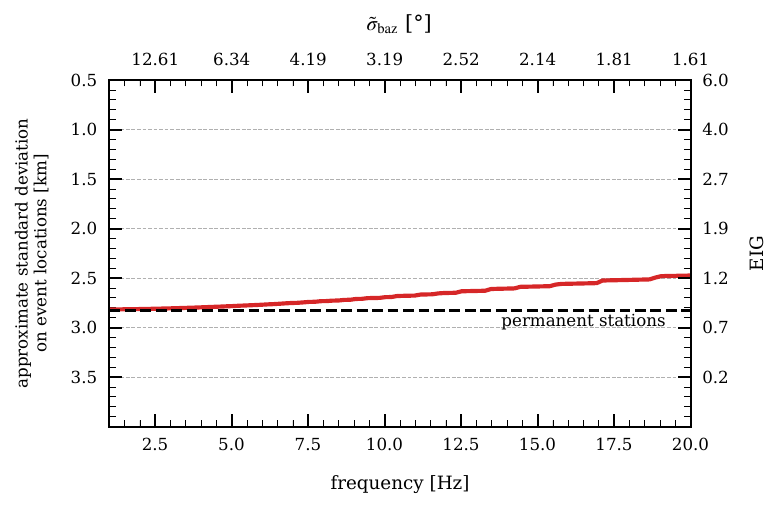}
  \caption{
    EIG and approximate posterior standard deviation for the Endurance site (red) as a function of frequency for the NYMA array. The EIG is calculated using the nested Monte Carlo estimate (Equation \eqref{eqn:EIG_MC_nested}) with $N=N_\mathrm{evi}=10,000$ and $N_\mathrm{like}=500$. The y-axis limits are intentionaly set to the same values as in Figure \ref{fig:02_array_eig_all_LA_std_baz} to allow for a direct comparison.
  }
  \label{fig:02_array_eig_std_baz}
\end{figure}

Using this approximate uncertainty for array measurement, we can calculate the EIG of the current UK seismic network combined with the NYMA array. We calculate the EIG as a function of frequency using the same prior distribution as in section \ref{subsec:noise_boulby} for seismic stations and the uncertainty derived above for array measurements. We set the magnitude range to ${M}=[1, 3]$ as this represents where the current network detects some but not all events. To focus on azimuthal data alone, we exclude potential travel time data with higher signal-to-noise ratios that the array might provide. Due to its close proximity, the Boulby site results offer insights into how this additional information would affect the EIG.

Figure \ref{fig:02_array_eig_std_baz} displays these results with y-axis limits matching Figure \ref{fig:02_array_eig_all_LA_std_baz} (see below) to facilitate direct comparison. For more intuitive understanding of frequency dependence, we show the approximate backazimuth uncertainty for an inclination of 25$^\circ$ and a subsurface velocity of 3000~m/s. This inclination corresponds to the expected angle of the incoming wavefield for a source at approximately 1~km depth and 200~km distance.

The information gain from adding an array to the current network remains limited, even for backazimuth uncertainties below 2$^\circ$. Array data alone provide insufficient additional information to the existing UK seismic station network. This limitation exists because the azimuthal data lacks sufficient accuracy to constrain event locations better than the prior distribution and the information already contained in travel time data. Even for events detected by the array but not by the current network, array data alone cannot provide substantial information gain compared to a low-noise onshore station or a single OBS station.

\begin{figure}
  \centering
  \includegraphics[width=0.5\textwidth]{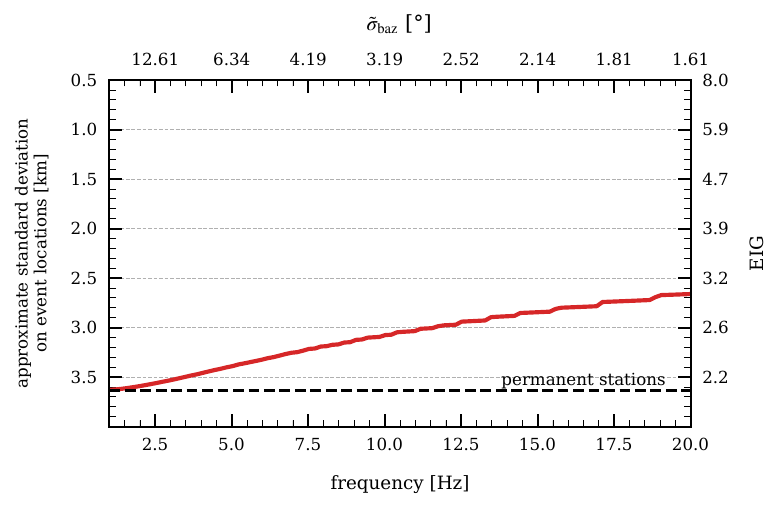}
  \caption{
    EIG and approximate posterior standard deviation for all licence areas as a function of frequency for the NYMA array. The EIG is calculated using the nested Monte Carlo estimate (Equation \eqref{eqn:EIG_MC_nested}) with $N=N_\mathrm{evi}=10,000$ and $N_\mathrm{like}=500$. Approximate backazimuth uncertainty is shown for an inclination of 25$^\circ$ and a subsurface velocity of 3000~m/s. The y-axis limits are intentionally set to the same values as in Figure \ref{fig:02_array_eig_std_baz} to allow for a direct comparison.
  }
  \label{fig:02_array_eig_all_LA_std_baz}
\end{figure}
When considering a substantially more spread out prior distribution across all licence areas, the array data provides greater information gain compared to the Endurance area alone. We tested this using the same prior distribution as in section \ref{subsubsec:mul_obs_all_licences}, which defines a uniform distribution within the bounds of all potential SCS sites. The results are shown in Figure \ref{fig:02_array_eig_all_LA_std_baz}.

The current network plus the array performs similarly to having a single OBS station (ignoring the current network) if a backazimuth uncertainty below 2$^\circ$ can be achieved. The EIG exceeds that of the Endurance area alone because the prior covers a larger azimuthal range.

Although more accurate measurements than our Gaussian fit might be possible in practice, the results demonstrate that information gain remains limited even with very sharp array response functions. The Gaussian fit may be overly conservative; however, we have ignored other sources of uncertainty in the array data, such as velocity model uncertainty, side lobe selection issues, and variability in the array response function itself. This suggests our analysis likely overestimates rather than underestimates the information gain from array data.

\end{document}